\documentstyle[aps,twocolumn]{revtex}
\input amssym.def
\input amssym
\font\script=eusm10

\newcommand{\SC}{\mbox{\script C}}

\begin{document}
\title{ {\large \bf Quantum Mechanics and Algorithmic Randomness}}
\author{Ulvi Yurtsever}
\address{Jet Propulsion Laboratory, California Institute of
Technology, 4800 Oak Grove Drive, Pasadena, California 91109-8099.}
\date{January 31, 1999\thanks{Revised version. To appear in {\sl Complexity},
November-December 2000.}}
\maketitle
\begin{abstract}
A long sequence of tosses of a classical coin produces an apparently
random bit string, but classical randomness is an illusion:
the algorithmic information content of a classically-generated bit string
lies almost entirely in the description of initial conditions.
This paper presents a simple argument that, by contrast, a sequence of
bits produced by tossing a quantum coin is, almost certainly,
genuinely (algorithmically) random.
This result can be interpreted as a
strengthening of Bell's no-hidden-variables theorem, and
relies on causality and quantum entanglement in
a manner similar to Bell's original argument.

~~~~~

\end{abstract}
\parskip 4pt

~~~~~~

\noindent{\bf 1. Notions of randomness}

As a mathematical concept, randomness (or, more precisely, the
notion of a random string of bits) has a convoluted history, with
various attempts at formalization going through multiple series of
refinements prompted by multiple setbacks (see [1], Chap.\ 1, for a
guide to these developments). Current thinking on randomness encompasses
two broadly-useful notions: pseudo-randomness and algorithmic
randomness. Random bit sequences of the first kind, pseudo-random
sequences, are derived from a small seed (initial data) through a
deterministic evolution algorithm, and pass as many practical
statistical tests of randomness as desired. Such sequences locally
``behave as if they are truly random," and are frequently useful in
statistical (e.g.\ Monte Carlo) simulations. A stronger variant of
pseudo-randomness, the notion of a cryptographically-secure
pseudo-random sequence, is motivated by applications involving secure
communication channels. To be useful in cryptography, a pseudo-random
sequence must have a generating algorithm which is secure from
eavesdroppers, in the sense that an adversary must be unable to deduce
the seed (in polynomial time) by observing a small part of the sequence
the seed generates ([2]). However, the strongest precise notion of
randomness so far developed is that of algorithmic incompressibility.
The central concept that underlies this notion is the Kolmogorov (or
algorithmic) complexity of a binary string, defined as the length (in
bits) of the minimal program that would output the given string when run
on a fixed universal Turing machine. A (long) bit string of length $N$ is
algorithmically random, or ``patternless," if its Kolmogorov complexity
is greater than $N$; if its complexity is significantly less than $N$,
then the string contains patterns which may be exploited to build a
description (algorithm) for it shorter than its length, i.e., the string
is algorithmically compressible.

An algorithmically random string is not only statistically
random (and, in fact, cryptographically secure), but it also captures
the intuitive idea of a truly random sequence which underlies the
foundations of probability theory (allusions to strict randomness, which
are implicit whenever the probability theorist talks about selecting an
object ``at random" out of a collection of objects, can only be
formalized through the notion of an infinite sequence of truly random
bits). It is not difficult to show (see [1], Chapter 3) that an
algorithmically random bit string is guaranteed to pass all computable
statistical tests of randomness. In particular, as the length $N$ of the
string approaches infinity, any computable pattern of $k$ (not
necessarily contiguous) bits has an asymptotic frequency of occurrence
$2^{-k}$; in other words, there are no correlations between any set of
$k$ bits whose relative positions are computably determined. Moreover,
algorithmic randomness is a ubiquitous property of long strings: among
all possible infinite binary strings those which are algorithmically
random form a subset of full measure; in other words, an infinite string
chosen ``without bias" is almost certain to be algorithmically random.

Despite these formal strengths (or, more precisely, because of
them), the notion of algorithmic randomness has one major weakness: no
algorithmically incompressible binary string can ever be {\em
constructed} via a finitely-prescribed procedure (since, otherwise, such
a procedure would present an obvious algorithm to compress the string
thus obtained). Furthermore, given a binary string which is, in fact,
algorithmically random, it is impossible to prove this fact starting
from any set of axioms whose information content is smaller than the
length of the string itself ([1], Section 2.\,7). Hence, for long enough
strings, algorithmic randomness is unprovable in any familiar
mathematical system based on set theory and logic (a fact closely
related to G{\"o}del's incompleteness theorems; see [3] for a discussion
of this connection). This ``fundamentally non-constructible" nature of
algorithmic randomness raises the following question: Is there any physical
process which naturally gives rise to an incompressible string of
binary outcomes where both the initial conditions and the dynamics are
finitely prescribed and precisely controlled?

Indeed, it turns out that quantum mechanics naturally provides
examples of such processes. It is a natural
question to ask whether quantum
mechanical probabilities are ``genuine;" by the
discussion above, showing that quantum probabilities are genuine
is equivalent to demonstrating that quantum
measurement outcomes are truly (algorithmically) random.
Such a demonstration is the basic claim of this paper.

\noindent{\bf 2. Randomness in classical and quantum physics}

A long string of pseudo-random bits produced by computer
passes all practical statistical tests of randomness (provided the
algorithm used is sound, see [4]), but it is not truly (algorithmically)
random: the information content of the string (its Kolmogorov
complexity) is bounded by the size of the generating algorithm plus a
few extra bits which specify the random number seed, typically a much
smaller quantity overall than the string's (arbitrarily large) length.
Similarly, provided it is continued to long enough lengths, a random
string of bits (such as a sequence of coin tosses) produced by any
classical physical system---of which a digital computer, or a Turing
machine, is a special example---is not truly random: its complexity is
bounded by the size of its ``algorithm," i.e.\ the deterministic
physical laws which govern its evolution, plus the size of a description
of the initial conditions on which those laws act. If the sequence of
bits is continued much longer than the size of this combined
description, the length of the resulting bit string is much larger than
its algorithmic complexity, which implies that the string is non-random
(compressible) even though, again, all statistical randomness tests may
be satisfied (as they would be if, e.g., the underlying dynamical system
is chaotic; \nolinebreak see [5]). The algorithmic information content
of a classically-produced bit string is contained entirely in the
description of initial conditions (with a small additional contribution
from the dynamical laws of evolution). 

Are there any physical systems that can generate arbitrarily
long, truly random (incompressible) bit strings starting from an initial
state with a simple description?
Note that the ``simplicity" of the initial
conditions is crucial:
Clearly, in classical physics one can start from initial conditions of
high algorithmic complexity and generate data strings of high
complexity (e.g. as described in [5]),
but then one gets only as much complexity as one is willing
to pay for up front, and no more. Even though ``almost all" initial
conditions in classical physics are algorithmically random, what is
required here is to start from a (simple) initial state which is
precisely controlled, and generate measurements
of unbounded complexity from it via its dynamical evolution.
Indeed, in this paper I will present an
argument that, if violations of relativistic causality are to be ruled
out, a bit string obtained as a result of binary measurements performed
on a string of identical copies of the same quantum state (where the
measurements yield 0 vs.\ 1 with equal probability) {\em must} be almost
surely (i.e.\ with probability that approaches 1 as the length of the
string grows to infinity) incompressible. More generally, my argument
shows that when the binary quantum measurements yield 1 with some
(non-trivial) probability $p$, the resulting bit string has (almost
surely) the maximal algorithmic complexity consistent with that
probability $p$. Unlike a classical system, the information ``contained"
in a quantum state cannot be compactly encoded in a description of
initial conditions. Note that such complete encoding of a quantum
system's algorithmic complexity in its initial conditions would be
possible if quantum mechanics were equivalent to a classical theory with
microscopic, local hidden variables; therefore the argument in this
paper can be further interpreted as a modest strengthening of Bell's
no-hidden-variables theorem ([6]). In fact, the argument here relies on
causality (i.e.\ locality) and quantum entanglement in a manner similar
to Bell's original argument.

Why is local causality necessary to infer such a basic fact as
the true randomness of quantum probabilities? Consider standard quantum
theory, consisting of the complex Hilbert space structure, linearity
(superposition), unitary evolution, and the standard theory of
measurement with the probability interpretation of inner-products. The
incompressibility of binary-measurement outcomes cannot be deduced from
these principles alone, since a computer-simulated quantum world, where
probabilistic outcomes are achieved through the use of random-number
generators, will respect these principles (and their observable
statistical consequences) just as well: The algorithmic complexity of a
string of quantum measurement outcomes is {\em not} determined simply by
the statistics of those outcomes, which statistics are
governed by the density matrix, or wave function, of the system, which
is in turn governed by the physical laws encoded in the principles
outlined above. For example, the binary expansion of the Champernowne
number ${\cal C} = .0100011011001\ldots$
is statistically random ([7]), but it is not algorithmically random. If
the outcomes from binary measurements on a quantum state were identical
to a segment of the binary expansion of ${\cal C}$
(with some arbitrary starting point), this would not violate any of the
statistical predictions of quantum mechanics. If this were the case
repeatedly, or, more precisely, if quantum measurement outcomes were (with
some finite probability) algorithmically compressible (but still
statistically random) just like the binary expansion of ${\cal C}$, then the
result claimed in this paper would be violated ([8]).

A more dramatic way to highlight this point is the following:
Suppose we simulate quantum mechanics and quantum-measurement outcomes
on a digital computer, using a very high-quality random-number
generator. The statistical structure of our simulated quantum world
would be identical (to arbitrarily high accuracy) with that of the real
world, but all of our bit strings from successively-run measurements
would be highly compressible. How do we know, then, that there is not some
universal random-number generator which generates the probabilistic
outcomes of quantum measurements (with the right statistics) in
the real world? It is easy to imagine random-number generators with
large-enough seeds (with seeds, say, of order 1-megabyte in size) such
that it is essentially impossible to discover their existence by
experiment (for example, to make sure
that we discover---via discovering repetitive
occurrences of the same string---that the bit strings from our
laboratory quantum measurements are generated by a hidden random-number
generator with 1-megabyte seed, it is necessary that we observe
as many as $2^{2^{23}}+1$ measurement sequences).

The result presented in this paper shows that
if violations of local causality are to be ruled out there can be no
such universal random-number generators in the real quantum world, and that,
as a result, it is not possible to simulate quantum mechanics on a
digital computer (i.e., a Turing machine);
quantum randomness is ``uncomputable"
in this sense. This fundamental lack of computability
of quantum phenomena may have certain far-reaching implications;
for example, if quantum-mechanical processes play a significant role in
the activities of biological neural systems, then brain activity cannot
be simulated faithfully on a digital computer, no matter how elaborate
the simulation.

Let us turn now to the presentation of the main argument:

\noindent{\bf 3. Entangled spin pairs and spacelike communication}

Consider a pair of spin-$\frac{1}{2}$ particles in the singlet (zero
total spin) quantum state
\begin{equation}
\left|\psi \right> = \frac{1}{\sqrt{2}}(\left| {\uparrow}_1 \right> 
\otimes \left| {\downarrow}_2 \right> 
-\left| {\downarrow}_1 \right> \otimes \left| {\uparrow}_2 \right> ) \; ,
\end{equation}
where, with respect to spin
measurements along an arbitrary direction axis,
$\left| {\uparrow}_i \right>$ and $\left| {\downarrow}_i \right>$ denote
the standard (orthonormal) eigenstates (spin-up and spin-down)
for spin $i$ ($i=1, \; 2$). Consider 
a long stream of such pairs produced
by some (stationary) source, all pairs created in exactly the same entangled
quantum state $\left|\psi \right>$ given by Eq.\,(1), and each
particle in the pair flying away from the source in opposite
directions. I will assume that two observers, Bob and Alice,
are positioned to perform observations at the opposite ends of
this pair of particle beams, where their measurements are performed
only after the particles have flown apart across a
spatial distance so large that any pair of observations
at the respective ends of the beam during the observers' lifetime
are spacelike-separated events in (flat) spacetime. It has always been
a useful question to ask whether the observers can make use of the
correlations inherent in the entangled state $\left|\psi \right>$
to transmit information to each other,
thereby violating relativistic causality.
This paper is no exception.

As is well known, standard laws of quantum mechanics
reveal that measurements performed on a single
pair of spins in the quantum state Eq.\,(1) can never be used to transmit
information between spacelike-separated observers ([9]), i.e., local causality
does not teach us anything new about quantum mechanics in this case. However,
a long stream of identical copies of the same state provide
significantly greater opportunities for communication, and it turns out
that imposing the no-spacelike-communications requirement here
leads to new knowledge on the structure of outcomes from
a string of quantum measurements.

To examine this structure, assume that Bob and Alice have agreed
beforehand (at some point in the distant past when they were
in causal contact) on a common axis (e.g.\ one which points
towards some distant quasar), and to measure
each spin arriving at their end in the
orthonormal bases $\{ \left| {\uparrow}_i \right> , \,
\left| {\downarrow}_i \right> \}$ along that axis (where $i=1$ for Bob
and $i=2$ for Alice). Upon performing a measurement, Bob records
his result as a 1-bit if the measured spin is in the
up direction (along $\left| {\uparrow}_1 \right>$) and as a 0-bit
otherwise, and Alice records her result as a 1-bit if
her measured spin is in the down direction (along
$\left| {\downarrow}_2 \right>$) and as a 0-bit otherwise. The nature
of the singlet state Eq.\,(1) guarantees that, as long as both
observers keep their measurements along the predetermined axis,
the bit strings Bob and Alice obtain at each end of the singlet-pair
stream are identical. The  observers can now
attempt to manipulate these two bit strings to build a
spacelike-separated communications channel.

The simplest strategy for communication Bob and Alice might think of
involves varying the frequency of 1's and 0's observed at
one end by varying the measurement procedure at the other. For example,
Alice and Bob know from standard quantum theory that as
long as they both follow the above measurement procedure
(which they previously agreed on),
their measured bit strings would each contain
asymptotically equal numbers of 1-bits and 0-bits,
reflecting a probability of $\frac{1}{2}$ for either
outcome. Now Bob can attempt to transmit a single bit of information to Alice
in the following way: to send a 0-bit, Bob does nothing special, preserving
the probabilistic structure of Alice's string the way she expects it;
to send a 1-bit, Bob changes his spin-measurement direction to point
at some new direction $( \theta , \phi )$
away from the original axis, and he keeps this
modified axis during a large (predetermined) number of measurements,
reverting back to the original axis only at the end of his bit-transmission
period. Can Alice reliably detect the transmission of this single bit
of information from Bob by examining the
minute changes (if any) in the
probability distribution of 0's and 1's in her bit string?
It is easy to see that the answer is no: although by changing
his spin-measurement direction Bob will cause Alice's bit string
to be different
from his (in case Bob decides to send a 1-bit), there is no way for
Alice to reliably discover this difference (other than a
direct comparison with
Bob's string, copied via a conventional communication channel); no
matter which bit Bob decides to send, Alice's
string has exactly the same probability distribution of 1's
and 0's (namely, probability precisely $\frac{1}{2}$ for each outcome).

Although the proof of this result is relatively easy for the singlet
state Eq.\,(1), it will be useful to briefly review it for the more
general case (where it is considerably less obvious) in which
the spins in the pair stream are in a generic
entangled quantum state $\left|\psi \right>$ given, instead of Eq.\,(1), by
\begin{equation}
\left|\psi \right> = \alpha \left| {\uparrow}_1 \right>
\otimes \left| {\downarrow}_2 \right>
+ \beta \left| {\downarrow}_1 \right> \otimes \left| {\uparrow}_2 \right> \; ,
\end{equation}
where $\alpha$ and $\beta$ are complex numbers satisfying the normalization
condition
\begin{equation}
|\alpha |^2 + |\beta |^2 = 1 \;
\end{equation}
with the singlet state corresponding to the
special case $\alpha=-\beta=1/\sqrt{2}$.
From Bob's point of view, $\left|\psi \right>$ corresponds to
a pure state $\rho= \left| \psi \right> \left< \psi \right|$ which,
when averaged (``traced") over all possible spin states of particle 2,
reduces to an effective density matrix
\begin{eqnarray}
{\rm Tr}_2 \; \rho & = & \left< {\uparrow}_2 \right| \rho \left|
{\uparrow}_2 \right> + \left< {\downarrow}_2 \right| \rho \left|  
{\downarrow}_2 \right> \; \nonumber \\
& = & |\alpha |^2 \left| {\uparrow}_1 \right> \left<  
{\uparrow}_1 \right| + |\beta |^2 \left| {\downarrow}_1 \right> \left<  
{\downarrow}_1 \right| \nonumber \\
& = &
\left(
    \begin{array}{cc}
      |\alpha |^2 & 0 \\
      0 & |\beta |^2 
    \end{array}
  \right) \; 
\end{eqnarray}
living in the (spin) Hilbert space of particle 1.
As long as Bob and Alice both follow the measurement
procedure they agreed on (i.e.\ as long as they measure repeatedly
along the direction axis which defines the spin bases $\{ \left|
{\uparrow}_1 \right> , \; \left| {\downarrow}_1 \right> \}$
and $\{ \left|
{\uparrow}_2 \right> , \; \left| {\downarrow}_2 \right> \}$),
it is clear from Eqs.\,(2) and
(4) that Bob's probability of observing a 1-bit
(spin up) is $|\alpha |^2$, which is exactly equal to Alice's probability
of observing a 1-bit (spin down). In fact, just as in the special
case of the singlet [Eq.\,(1)], so also here the bit strings Alice
and Bob obtain under the standard measurement procedure are identical.
Now suppose that Bob, in his attempt to transmit a 1-bit to Alice,
modifies his spin-measurement axis to point in
a new direction along which the
spin-up eigenstate is given by $\left| {\nearrow}_1 \right>
=c \left| {\uparrow}_1 \right> + d \left| {\downarrow}_1 \right>$, where
$c$ and $d$ are complex numbers with $|c|^2+|d|^2=1$. The new \{spin-up,
spin-down\} eigenbasis (for the Hilbert space of particle
1) along this modified direction is then given by the
orthonormal state vectors
\begin{eqnarray}
\left| u \right> & \equiv & \left| {\nearrow}_1 \right> \; \, = \, \;
c \left| {\uparrow}_1 \right> + d \left| {\downarrow}_1 \right> \; ,
\nonumber \\
\left| v \right> & \equiv & \left| {\swarrow}_1 \right> \; \, = \, \; 
- \overline{d} \left| {\uparrow}_1 \right> + 
\overline{c} \left| {\downarrow}_1 \right> \; .
\end{eqnarray}
The probability for Bob to observe a 1-bit (spin-up) in this new
eigenbasis is the expectation value [with respect to the effective mixed
state Eq.\,(4)] of the projection operator
$ \left| u \right> \left< u \right|$ on the
eigenstate $\left| u \right> = \left| {\nearrow}_1 \right>$:
\newpage
\begin{eqnarray}
{\rm Prob}(1_1) & = & {\rm Tr}_1 \left[
\left(
    \begin{array}{cc}
      |\alpha |^2 & 0 \\
      0 & |\beta |^2
    \end{array}
  \right) \right. \nonumber \\
& & \left. 
\begin{array}{cc}
      ~ & ~ \\
      ~ & ~
    \end{array}
( c \left| {\uparrow}_1 \right> + d \left| {\downarrow}_1 \right> )
(\overline{c} \left< {\uparrow}_1 \right| + 
\overline{d} \left< {\downarrow}_1 \right| ) \right] \nonumber \\
& = & {\rm Tr} \left[
\left( 
    \begin{array}{cc}
      |\alpha |^2 & 0 \\
      0 & |\beta |^2
    \end{array}
  \right) \left( 
    \begin{array}{cc}
      |c |^2 & c \overline{d} \\
      d \overline{c} & |d|^2
    \end{array}
  \right) \right] \nonumber \\
& = &
|\alpha |^2 |c|^2 + |\beta |^2 |d|^2 \; , 
\end{eqnarray}
where the subscripts ``1" denote that the corresponding objects
are associated with
particle 1. Bob's new probability for observing a ``1-bit" is manifestly
different from the original probability $|\alpha |^2$,
and his modified bit-string
will reflect this difference in the new asymptotic distribution of 1's and 0's.
But even so, Alice still has no way of detecting the change which
Bob's decision to modify his axis has induced on {\em her} bit string:
Despite the drastic change in the statistics at Bob's side,
Alice will continue to observe a bit string where 1's
still have the original asymptotic frequency of
$|\alpha |^2$. Indeed, rewriting the entangled state Eq.\,(2) in the new
basis (for the Hilbert space of particle 1) Eq.\,(5) 
\begin{eqnarray}
\left|\psi \right> & = & \alpha \; ( \overline{c} \left| u \right>
- d \left| v \right> )  
\otimes \left| {\downarrow}_2 \right>
\; + \; \beta \; (\overline{d} \left| u \right>
+ c \left| v \right> ) 
\otimes \left| {\uparrow}_2 \right> \nonumber \\
& = & \left| u \right> \otimes ( \beta \overline{d}
\left| {\uparrow}_2 \right> +  \alpha \overline{c}
\left| {\downarrow}_2 \right> ) \nonumber \\
& + &
\left| v \right> \otimes ( \beta c 
\left| {\uparrow}_2 \right> -  \alpha d 
\left| {\downarrow}_2 \right> )  \; 
\end{eqnarray}
and making use of Eq.\,(6),
it is straightforward
to compute Alice's new probability of observing a ``1-bit" in her
bit string (recall: for Alice ``1-bit" $\equiv$ spin-down):
\begin{eqnarray}
{\rm Prob}(1_2) & = &
{\rm Prob}(1_1)
\frac{|\alpha \overline{c}|^2}{|\beta \overline{d}|^2 + |\alpha \overline{c}|^2}
\nonumber \\
& + & {\rm Prob}(0_1) 
\frac{| - \alpha d|^2}{|\beta c|^2+| - \alpha d|^2}
\nonumber \\
& = &
|\alpha c|^2
\; + \; (1 - |\alpha c|^2 - |\beta d|^2) 
\frac{|\alpha d|^2}{|\beta c|^2+|\alpha d|^2}
\nonumber \\
& = & |\alpha |^2 \; ,
\end{eqnarray}
where the final equality follows at once from the normalization condition
$c^2 + d^2 = 1$ after one notices the identity
$1 - |\alpha c|^2 - |\beta d|^2 = |\beta c|^2+|\alpha d|^2$
which follows from Eq.\,(3).

Realizing that their attempts at communication via
manipulating the statistics of each other's measurements
are doomed to fail, Bob and
Alice may turn, in desperation, to the only remaining structural
signature their bit strings have: algorithmic complexity ([10, 1]). The
algorithmic (or Kolmogorov) complexity of a bit string $S$ is the length
(in bits) of the shortest program that would output $S$ when run on a
fixed Universal Turing Machine (UTM). When a string $S_n$ of length $n$
is algorithmically random (patternless), its Kolmogorov complexity is
comparable to $n$: $K(S_n ) \sim n$; if its complexity is significantly
less than $n$, then $S_n$ contains patterns which may be exploited to
build an algorithmic description for it shorter than its actual length,
i.e., such a string is compressible. In general, the quantity $K(S_n)$
is meaningful only in the limit of very long strings
($n \rightarrow \infty$), since only in this limit independence from a
specific choice of UTM is assured ([11]).
While an incompressible string necessarily
has the same asymptotic fraction of 0-bits as 1-bits (i.e.\
$\frac{1}{2})$, more generally, a bit string $S$ in which the asymptotic
frequency of 1's is $p$ has an algorithmic complexity of at most
\begin{eqnarray}
K(S_n ) & \sim & n H(p) \; , \nonumber \\
H(p) & \equiv &
-p \log_2 p - (1-p) \log_2 (1-p) \; ,
\end{eqnarray}
where $S_n$ denotes
the first $n$ bits of $S$, and $H(p)$ is the Shannon entropy of the
probability $p$. A bit string which satisfies Eq.\,(9) has the maximal
algorithmic complexity (randomness) subject to the statistical
constraint imposed by the asymptotic 1-bit-frequency $p$. I will
refer to this property (or its negation) as $p$-incompressibility
(or $p$-compressibility) in what follows (the usual notions obtain when
$p=\frac{1}{2}$).

To avoid any confusion, it is perhaps appropriate here to
re-emphasize that the notions of algorithmic compressibility used in
this paper refer strictly
to finite (albeit very long) binary strings. Nevertheless,
the notions of incompressibility for infinite binary strings are
essentially the same notions (save for unavoidable formal differences). For practical
reasons, I will always use the finite-substring approach to incompressibility
of infinite binary strings as hinted in the previous paragraph
[Eq.\,(9)]. Accordingly, an infinite string $S$ is (by definition)
incompressible if and only if
\[
\lim_{n \rightarrow \infty} \frac{K(S_n )}{n} = 1 \; 
\]
(where $S_n$ denotes the finite substring consisting of the first $n$
bits of $S$), and it is $p$-incompressible if and only if
\[
\lim_{n \rightarrow \infty} \frac{K(S_n )}{n} = H(p) \; .
\]
For a discussion of other (equivalent) definitions of incompressibility
for infinite binary strings the reader should consult Ref.\,[1].

Let $p_N$ denote the probability that an $N$-bit-long segment
of Bob's (or Alice's) string of quantum measurements is $p$-compressible
[where $p=|\alpha |^2$ in the context of the discussion between
Eqs.\,(2) and (8)]. I will now sketch a proof that if the statement of
the main result of this paper [see the second paragraph in Sect.\,2
above] is false, in other words, if the probability $p_N$ is bounded
away from zero as $N \rightarrow \infty$, then a reliable (spacelike)
communications channel can be constructed through which Bob---using each
$N$-bit-long block as a carrier of one data bit---can send information to
Alice. Here is how the construction of this communications channel might
proceed:

First, Alice and Bob agree at the outset that Alice should
interpret any compressible $N$-bit block in her string as a 0-bit, and
any {\em in}compressible block as a 1-bit. Next, they agree on an
approximate value for the Universal Halting Probability $\Omega$
[computed with respect to a common choice of UTM;
see Eq.\,(10) in Sect.\,4 below], so that Alice can
determine, with a probability of error $p_\Omega$ less than 1, whether
or not a given segment of her bit string is compressible [this is
necessary because both the Kolmogorov complexity $K(\cdot )$ and
the map which takes $n$ to the $n$'th bit
of $\Omega$ are nonrecursive functions, therefore Alice cannot make her
determinations with absolute certainty; see Sect.\,4 below for
more details]. Now, to send a 0-bit to Alice, Bob does nothing (i.e.,
keeps his spin-measurement axis unchanged, pointing along the original
predetermined direction). To send a 1-bit, Bob ``scrambles" his
measurement sequence in the following way: first he generates a random
``template," a random bit string $T$ of length greater than $N$ which
is (almost surely) incompressible [Bob can obtain such a string, among
other ways, by using the evolution from random initial conditions of a
classical system (such as a roulette wheel) with greater than $N$ degrees of
freedom]. Then, Bob prepares a sequence of $N$ random measurement
directions using $T$ as a random-number generator, and performs
his next $N$ measurements along the successive orthonormal bases
associated with the successive directions from this random
sequence. This procedure of scrambling with the random
template $T$ guarantees that Bob's modified $N$-bit long string of
quantum measurements is almost surely $p$-incompressible [with $p=
\frac{1}{3}(|\alpha |^2 + 2 |\beta |^2)$, in the notation of
Eqs.\,(2)---(8)], and that Alice's corresponding string (which is now
different from Bob's) is also (almost surely) $p$-incompressible [with
$p=|\alpha |^2$, as in Eq.\,(8)].
The crucial requirement for Bob's choice
of $N$ measurements is that it should be guaranteed to be
free of any regularities that might be present in the $N$-bit
block (representing the 1-bit) he wishes to send Alice.
His scrambling procedure is designed to destroy any nascent
correlations that might exist in the original bit stream of
measurements, without introducing any extra correlations of
quantum-mechanical origin (hence his use of the classical roulette-wheel).
There are other ``scrambling"
procedures Bob might use to ensure this; the procedure just
described is only one natural choice.

The transmission protocol
described above establishes a noisy (asymmetric) binary communication
channel from Bob to Alice, connecting them across a spacelike spacetime
separation: Bob can now transmit a 1-bit to Alice (almost-)reliably,
and he can transmit a 0-bit unreliably.
For communication to be possible through this channel, the
channel capacity must remain nonzero in the limit $N \rightarrow
\infty$; Shannon's noisy-channel coding theorem ([12]) would then ensure
that coding schemes can be found which will allow transmission of
messages with arbitrarily small probability of error. Indeed, it is not
difficult to
show that the capacity of the channel just constructed remains
nonzero in the limit $N \rightarrow \infty$ if
it is assumed, contrary to the assertion of this paper, that the
probability $p_N$ remains bounded away from zero in this limit.

\noindent{\bf 4. Detailed analysis of the
entangled-spin communication channel}

I will denote the vanishing of a quantity $q$
in the limit $N \rightarrow \infty$
by the expression $q=O(\epsilon )$. For example, the argument of this paper
will show that to preserve relativistic causality
it is necessary to have
$p_N = O( \epsilon )$, where $p_N$ denotes the probability
that an $N$-bit-long segment
of Bob's (or Alice's) string of quantum measurements is $p$-compressible
[where $p=|\alpha |^2$ in the context of the discussion between
Eqs.\,(2) and (8) in Sect.\,3 above]. In the present discussion I will
not try to quantify the rate at which $p_N$ must decay to
zero; a more detailed and rigorous analysis,
to be presented separately in [13], is needed to provide
sharper estimates for the
asymptotic decay rate of $p_N$. However, for all probabilities
of order $O( \epsilon )$ to be discussed here, including for $p_N$,
the true decay rates can be shown to be exponential
(see [13] for details).

The ``Universal Halting Probability"
alluded to in Sect.\,3 above is defined by
\begin{equation}
\Omega \; = \sum_{\pi : \; \pi \; {\rm halts}} 2^{-l(\pi )} \; \; , 
\end{equation}
where the sum is over all prefix-free (i.e.\ no $\pi$ is the
prefix of another $\pi '$) programs $\pi$ which halt when run on a fixed
UTM, and $l(\pi )$ denotes the length of $\pi$ in bits
(convergence of the sum in Eq.\,(10) to a real number less
than 1 is ensured by Kraft's inequality; see Refs.\,[1] and [14]).
Thus defined,
$\Omega$ is the probability that a randomly chosen program will halt
when run on the given UTM (for further details see Refs.\,[10, 1]).
Complete knowledge of $\Omega$ would allow one to solve the halting
problem ([15]), consequently $\Omega$ cannot be recursively
evaluated (this is related to the fact that
the Kolmogorov complexity $K$ is a nonrecursive
function); moreover, $\Omega$ is
an incompressible real number (i.e.\ its binary expansion is an incompressible
bit string; for a lucid discussion
of these and other magical properties
of $\Omega$ consult [16]).

Returning now to the
construction described in the previous section (Sect.\,3), if
Alice wanted to decide {\em with certainty} whether or not
a given $N$-bit block in her string is $p$-compressible, she would
need to know $\Omega$ (or at least the first $N$ bits in the
binary expansion of $\Omega$) with absolute certainty ([17]).
However, since Alice and Bob's construction merely attempts to
create a {\em noisy} communications channel,
all Alice really needs to know is an approximation
to $\Omega$ such that she can make her decisions with a fixed probability of
error $p_\Omega < 1$ (see [13] for details). The error probabilities for
Alice are then
\begin{eqnarray}
{\rm Prob} ( & &\mbox{{\scriptsize Alice falsely decides a $p$-compressible
string}} \nonumber \\
& & \mbox{{\scriptsize to be $p$-incompressible}}) = p_\Omega \; ,
\end{eqnarray}
while
\begin{eqnarray}
{\rm Prob} ( & & \mbox{{\scriptsize Alice falsely decides a $p$-incompressible 
string}} \nonumber \\
& & \mbox{{\scriptsize to be $p$-compressible}}) = 0 \; .
\end{eqnarray}
Now, a binary communication channel with asymmetric
bit-error probabilities given by
\begin{eqnarray}
p_0 & \equiv &
{\rm Prob}( \mbox{{\scriptsize a 0-bit is flipped in transmission}}) \nonumber \\
& = & {\rm Prob}(1_{\rm out}
\mid 0_{\rm in}) \; , \nonumber \\
p_1 & \equiv &
{\rm Prob}( \mbox{{\scriptsize a 1-bit is flipped in transmission}}) \nonumber \\
& = & {\rm Prob}(0_{\rm out}
\mid 1_{\rm in}) \; 
\end{eqnarray}
can be shown to have a channel capacity (see [13] for a detailed
derivation)
\begin{eqnarray}
C(p_0 , p_1 ) & = & \log_2 \left[ 1+ 2^{r(p_0 , p_1 )} \right] \nonumber \\
& - & p_0 \, r(p_0 , p_1) - H(p_0 ) \; \; ,
\end{eqnarray}
where
\begin{equation}
r(p_0 , p_1 ) \equiv \frac{H(p_0 ) - H(p_1 )}{p_0 + p_1 -1} \; \; .
\end{equation}
For the binary channel which Bob and Alice would obtain with their
communication
protocol, it is easy ([13]) to show that
\begin{eqnarray}
p_0 & = & {\rm Prob}(1_{\rm out}
\mid 0_{\rm in}) \; = \; p_N p_\Omega + (1-p_N ) \nonumber \\
& = & 
1- (1-p_\Omega )p_N \; ,
\end{eqnarray}
while
\begin{eqnarray}
p_1 & = & {\rm Prob}(0_{\rm out}
\mid 1_{\rm in}) \; = \; 0\,[1-O(\epsilon )] + (1-p_\Omega )O(\epsilon )
\nonumber \\
& = & 
O(\epsilon ) \; . 
\end{eqnarray}
Inspection of Eqs.\,(14)---(15) reveals that,
in general, the binary asymmetric
channel capacity $C(p_0 , p_1 )$ vanishes along the diagonal
$\{ p_0 + p_1 = 1\}$, and is second order in the distance
$1-p_0 - p_1 $ away from this diagonal in its vicinity ([13]).
Consequently, substitution of Eqs.\,(16)---(17)
in Eqs.\,(14)---(15) makes it straightforward to verify the main claim
of this paper, namely: as long as $p_N > O(\epsilon )$, the (spacelike) channel
capacity from Bob to Alice remains bounded away from zero
in the limit $N \rightarrow \infty$ [in other words, $C_N > O(\epsilon )$
as long as $p_N > O(\epsilon )$].

\noindent{\bf 5. Conclusions}

The argument presented above proves that,
if relativistic causality is to be preserved,
a bit string generated by binary measurements
performed on a string of identical copies of
the quantum state Eq.\,(2) must be almost surely (i.e.\ with
probability that approaches 1 as the length of the string grows
to infinity) maximally algorithmically random. This result follows
directly from the most basic laws of standard quantum mechanics and
quantum measurement theory, and those laws do not grant any privileged
status to the specific entangled form of the state Eq.\,(2).
Given an arbitrary quantum state
$\left| \Psi \right>$, any binary measurement determines a
choice between two projections:
a projection either onto a special state, $\left| 1 \right>$, say,
or to the orthogonal complement $\left| 1 \right>^{\perp}$
of this state
in the Hilbert space of the system ([18]). But the true binary decision is
between the state $\left| 1 \right>$ and the projection of
$\left| \Psi \right>$ on $\left| 1 \right>^{\perp}$; denote
this projection by $\left| 0 \right>$, and we are back
in the two-(complex)-dimensional Hilbert-space geometry
(e.g.\ the geometry of the subspace spanned by
$\left| 1 \right> = \left| {\uparrow}_1 \right>
\otimes \left| {\downarrow}_2 \right>$ and
$\left| 0 \right> = \left| {\downarrow}_1 \right>
\otimes \left| {\uparrow}_2 \right>$) of Eq.\,(2)
and the discussion which follows it. Combined with the
general unitary invariance of quantum mechanics, this
argument shows that
my result on the algorithmic
randomness of binary-measurement outcomes
applies just as well to the arbitrary
state $\left| \Psi \right>$ as to the specific entangled state Eq.\,(2).

Furthermore, since one cannot build a physical system
which can make copies of an arbitrary ensemble
of quantum states (a quantum
``copier" can make duplicates of no more than
as many different states as
would fit within an orthogonal set,
as explained, e.g., in Ref.\,[19]), ``a string of identical copies" of
a given quantum state is a meaningful construction
only in the context of a physical
process which creates such copies in unlimited succession,
such as the process I discussed in Sect.\,3 above immediately following
Eq.\,(1). Nevertheless,
the argument from unitary equivalence described in the
previous paragraph can be
used once again to further enlarge the domain
of application of the present result,
namely: any string of binary quantum
measurements which can be mapped unitarily onto another
must give rise to a bit string
of outcomes with the same statistical and
algorithmic structure as the string it
is unitarily mapped onto. Consequently, given any
sequence of measurements unitarily equivalent to successive binary
measurements on a fixed state $\left| \Psi \right>$
the bit string of successive outcomes is subject to
the incompressibility result of this paper.

\begin{center}
\noindent{\bf Acknowledgement}
\end{center}

I would like to thank the reviewers for
pointing out several corrections
and suggestions for improvement.

\newpage

\begin{center}
{\bf NOTES AND REFERENCES}
\end{center}

\noindent{\bf 1.} M.\ Li and P.\ Vitanyi, {\it An Introduction to
Kolmogorov Complexity and its Applications} (Springer-Verlag,
New York 1993).

\noindent{\bf 2.} A.\ Shamir, {\it On the generation of
cryptographically strong pseudo-random sequences} in {\it Proc.\
ICALP} (Springer, 1981). Also see M.\ Bellare and S.\ Goldwasser,
{\it Lecture Notes on Cryptography}, available on the world-wide-web at
{\tt http://www-cse.ucsd.edu /users/mihir/papers/gb.html}.

\noindent{\bf 3.} G.\ J.\ Chaitin, Int.\ Jour.\ Th.\ Phys.\ {\bf 22},
941 (1982).

\noindent{\bf 4.} D.\ E.\ Knuth,
{\it The Art of Computer Programming.\ Vol.\ 2:\ Seminumerical
Algorithms} (Addison Wesley Longman, Reading, Massachusetts 1998).

\noindent{\bf 5.} A trivial exception to this would be a classical
chaotic system with initial conditions given (to arbitrarily high
precision) by real numbers which are themselves incompressible
(algorithmically random). Because of the exponential loss of accuracy in
its chaotic evolution, to describe an $N$-bit string of ``coin tosses"
in this case one would need to know the initial conditions to an
accuracy of approximately a part in $2^N$ (an accuracy of $N$ binary
digits), which implies an algorithmic complexity of order $N$ for the
resulting bit string. Obviously, such a string {\em can} be
algorithmically random (incompressible). Note, however, that the
initial conditions do not have a finite (or short)
description in this case.

\noindent{\bf 6.} J.\ S.\ Bell, Physics {\bf 1},
195 (1964). Also reprinted in {\it Speakable
and Unspeakable in Quantum Mechanics} (Cambridge University Press,
Cambridge 1987).

\noindent{\bf 7.} The Champernowne number {\SC}
is defined to be the real number in $[0,1]$ whose binary expansion
is given by the Champernowne sequence
$0100011011000001010011100101110111\ldots$,
a provably statistically random
bit string which is not algorithmically random, where
the lexicographically ordered list of all strings of length $k$ are followed
by the list of all those of length $k+1$ as $k$ ranges from $1$ to $\infty$.

\noindent{\bf 8.} Of course, one can always add the algorithmic
incompressibility of successive measurement outcomes
as a basic principle (an ``axiom") to standard quantum
mechanics without contradicting observation in any way; this would
presumably allow one to deduce causality as a theorem. From this
viewpoint, the result claimed here suggests that algorithmic
incompressibility and relativistic causality are interchangeable as
principles of quantum mechanics. Nevertheless, local causality is
arguably more natural than algorithmic incompressibility as a candidate
axiom for quantum (or, for that matter, any other physical) theory.

\noindent{\bf 9.} Measurements performed
on one member of an
entangled pair cannot alter the expectation values
of operators acting on the other; see, e.g., D.\ Bohm, {\it Quantum Theory}
(Prentice Hall, Englewood Cliffs 1951).

\noindent{\bf 10.} G.\ J.\ Chaitin, IBM Journal of Research and Development,
{\bf 21}, 350 (1977); Advances in Applied Mathematics
{\bf 8}, 119 (1987).

\noindent{\bf 11.} I will be somewhat cavalier in my quantitative treatment of
Kolmogorov complexity. For example, a proper treatment of the
complexity measure $K$ would define it in terms of
``prefix-free" (or, equivalently,
self-delimiting) UTM programs, and accordingly
replace Eq.\,(9)
with the more accurate
$K(S_n ) \sim n H(p) + 2 \log_2 n$ for a
maximally-random string $S$. Also, the symbol ``$\sim$" has a rather
precise meaning in the algorithmic-information-theory literature which
I will gloss over. These technical details are not essential to the flow
of my argument, and they can be filled in from the references [10] and [1],
especially the book by Li and Vitanyi; I will give a more detailed and
rigorous account of my analysis elsewhere ([13]).

\noindent{\bf 12.} C.\ E.\ Shannon, Bell Sys.\ Tech.\ Journal
{\bf 27}, 379; 623 (1948). See also C.\ E.\
Shannon and W.\ W.\ Weaver, {\it The Mathematical Theory of
Communication} (University of Illinois Press, Urbana 1949).

\noindent{\bf 13.} U.\ Yurtsever, manuscript in preparation.

\noindent{\bf 14.} T.\ M.\ Cover and J.\ A.\ Thomas, {\it Elements of
Information Theory} (Wiley-Interscience, New York 1991).

\noindent{\bf 15.} Given the exact value of $\Omega$ and a program
$\pi _0$ whose halting is to be decided,
begin running the given UTM
with all possible programs $\{ \pi \}$
arranged in a ``dove-tailed" input configuration,
and simply wait until either $\pi _0$
halts, or the sum Eq.\,(10) over all programs which already halted
accumulates to a value greater than $\Omega - 2^{-l(\pi _0 )}$
(which, when it happens, will guarantee that $\pi _0$ does not halt).

\noindent{\bf 16.} M.\ Gardner, Scientific American {\bf 241}, 20 (1979).

\noindent{\bf 17.} To make her decision about a bit string
$S$, Alice would simply run all
possible ``short" programs in the same manner as described above (in [15]),
wait until she is sure every program which will ever halt has
already done so
[by monitoring the accumulating sum Eq.\,(10) until it comes
close enough to her value of $\Omega$], and finally see if the string $S$ is
contained among the outputs of the halted programs (if it is, then
$S$ is compressible; otherwise, $S$ is incompressible).

\noindent{\bf 18.} Although this argument assumes $\left| \Psi \right>$
is a pure state,
it is not difficult to generalize its conclusion to
more general mixed states via reduction
to the pure-state case; see Ref.\,[13].

\noindent{\bf 19.} C.\ M.\ Caves and
C.\ A.\ Fuchs, {\it Quantum Information:
How Much Information in a State Vector?} quant-ph/9601025 (1996).

\end{document}